\documentclass[aps,prl,reprint,superscriptaddress,nobibnotes]{revtex4-1}
\usepackage{graphicx}
\usepackage{amsmath}
\usepackage{amssymb}
\usepackage[labeled,resetlabels]{multibib}
\newcites{S}{Supplemental References}
\usepackage{hyperref}

\newcounter{mysection}
\makeatletter
     \@addtoreset{figure}{mysection}
\makeatother

\makeatletter
\newcommand*{\balancecolsandclearpage}{%
  \close@column@grid
  \clearpage
  \twocolumngrid
}
\makeatother

\begin{document}
\title{Part-per-million quantization and current-induced breakdown \\of the quantum anomalous Hall effect}

\author{E.~J.~Fox}
\affiliation{Department of Physics, Stanford University, Stanford, CA 94305, USA}
\affiliation{Stanford Institute for Materials and Energy Sciences, SLAC National Accelerator Laboratory, 2575 Sand Hill Road, Menlo Park, California 94025, USA}

\author{I.~T.~Rosen}
\affiliation{Department of Applied Physics, Stanford University, Stanford, CA 94305, USA}
\affiliation{Stanford Institute for Materials and Energy Sciences, SLAC National Accelerator Laboratory, 2575 Sand Hill Road, Menlo Park, California 94025, USA}

\author{Yanfei Yang}
\affiliation{National Institute of Standards and Technology (NIST), Gaithersburg, MD 20899-8171, USA}

\author{George R.~Jones}
\affiliation{National Institute of Standards and Technology (NIST), Gaithersburg, MD 20899-8171, USA}

\author{Randolph E.~Elmquist}
\affiliation{National Institute of Standards and Technology (NIST), Gaithersburg, MD 20899-8171, USA}

\author{Xufeng Kou}
\affiliation{Department of Electrical Engineering, University of California, Los Angeles, CA 90095, USA}
\affiliation{School of Information Science and Technology, ShanghaiTech University 201210, China}

\author{Lei Pan}
\affiliation{Department of Electrical Engineering, University of California, Los Angeles, CA 90095, USA}

\author{Kang L.\ Wang}
\affiliation{Department of Electrical Engineering, University of California, Los Angeles, CA 90095, USA}

\author{D.\ Goldhaber-Gordon}
\email[To whom correspondence should be addressed; \\Email: ]{goldhaber-gordon@stanford.edu}
\affiliation{Department of Physics, Stanford University, Stanford, CA 94305, USA}
\affiliation{Stanford Institute for Materials and Energy Sciences, SLAC National Accelerator Laboratory, 2575 Sand Hill Road, Menlo Park, California 94025, USA}

\date{\today}

\begin{abstract}
In the quantum anomalous Hall effect, quantized Hall resistance and vanishing longitudinal resistivity are predicted to result from the presence of dissipationless, chiral edge states and an insulating 2D bulk, without requiring an external magnetic field. Here, we explore the potential of this effect in magnetic topological insulator thin films for metrological applications. Using a cryogenic current comparator system, we measure quantization of the Hall resistance to within one part per million and longitudinal resistivity under 10~m$\Omega$ per square at zero magnetic field. Increasing the current density past a critical value leads to a breakdown of the quantized, low-dissipation state, which we attribute to electron heating in bulk current flow. We further investigate the pre-breakdown regime by measuring transport dependence on temperature, current, and geometry, and find evidence for bulk dissipation, including thermal activation and possible variable-range hopping.
\end{abstract}

\maketitle

When doped with certain transition metals, chal\-co\-gen\-ide-based 3D topological insulators (TIs) can be made ferromagnetic, breaking time-reversal symmetry and opening a gap in the Dirac spectrum of the topological surface states~\cite{Yu2010,Chen2010,Checkelsky2012}. However, this gap should close where the component of the magnetization normal to the surface changes direction. In a thin-film sample uniformly magnetized in the out-of-plane direction, this transition occurs at the physical edge of the film as the surface normal switches direction in going from the top surface to the bottom. In the idealized theoretical picture of these systems, the 2D bulk is completely gapped, and one-dimensional channels arising at such boundaries are chiral and dissipationless due to the absence of available states for backscattering. The resulting edge conduction, which does not require an external magnetic field, is known as the quantum anomalous Hall (QAH) effect, and transport measurements are predicted to find vanishing longitudinal resistivity $\rho_{xx} = 0$ accompanied by quantized Hall resistivity $\rho_{yx} = \pm h/e^2$, where $h$ is Planck's constant and $e$ the electron charge.

Experiments have indeed found $\rho_{yx}\approx h/e^2$, but have not clearly demonstrated an insulating bulk as predicted. The first reported observation of QAH in Cr-doped (Bi,Sb)$_2$Te$_3$ at zero field measured $\rho_{yx}$ within a few percent of $h/e^2$, yet $\rho_{xx} \approx 2.5$ k$\Omega$~\cite{Chang2013}, far above the expected value. Subsequent works on the same material system~\cite{Checkelsky2014,Kou2014,Bestwick2015,Kandala2015,Mogi2015} and a V-doped analogue~\cite{Chang2015,Grauer2015} have replicated the effect, progressively improving on the degree of quantization and reducing longitudinal resistivity, but in all cases have found nonvanishing $\rho_{xx}$, indicating dissipative transport. Proposed explanations for this dissipation include thermally-activated bulk or surface carriers~\cite{Bestwick2015}, variable-range hopping (VRH)~\cite{Chang2013,Kawamura2017}, or the presence of extra, non-chiral edge states~\cite{Wang2013,Kou2014,Chang2015-2}. Moreover, $\rho_{yx}$ deviates from quantization and $\rho_{xx}$ rises rapidly with temperature, typically on a scale of hundreds of millikelvin~\cite{Bestwick2015,Kawamura2017,Chang2015-2,Liu2016}, unexpectedly small compared to the Curie temperature, which is generally tens of kelvin~\cite{Checkelsky2012,Chang2013,Checkelsky2014,Kou2014,Chang2015,Kandala2015,Mogi2015,Grauer2015,Lee2015,Li2016}. Angle-resolved photoemission spectroscopy results~\cite{Li2016} suggest that in V-doped films, this discrepancy may be due to the bulk valence band overlapping the surface state gap, while scanning tunneling microscopy~\cite{Lee2015} on Cr-doped bulk crystals shows strong spatial variations of the exchange-induced gap which could result in a small temperature scale in transport~\cite{Yue2016} despite an average gap size of $\sim$30 meV. However, a comprehensive explanation of the unexpected and non-ideal behavior in QAH is still lacking.

To investigate the degree to which dissipation can be removed, several studies~\cite{Bestwick2015,Chang2015,Liu2016} previously attempted to characterize samples with apparently well-quantized Hall resistance at temperatures of tens of millikelvin. The most precise of these measurements show $\rho_{yx} = h/e^2$ to within one part in $10^4$ and $\rho_{xx}$ as low as 1 $\Omega$ per square~\cite{Bestwick2015}. In contrast, measurements of the quantum Hall (QH) effect, in which similar vanishing $\rho_{xx}$ and quantized $\rho_{yx} = h/\nu e^2$ for integer $\nu$ is predicted due to Landau level formation, have found quantization of the Hall resistance to within a part in $10^9$ and the lowest longitudinal resistivity ever measured in a non-superconducting sample~\cite{Jeckelmann2001}. Because of the precision and reproducibility of such measurements, a conventional value of the von Klitzing constant $R_K = h/e^2$ is the basis for practical metrology of the ohm, even though maintaining a QH resistance standard~\cite{Poirier2009} requires very low cryogenic temperatures and large magnetic fields. The situation is however improving; in graphene, resistance quantization to within one part in $10^9$ can be measured at 5 K in a 5 T field~\cite{Ribeiro-Palau2015}.

Realization of the QAH effect raises the prospect of a future quantum resistance standard without need for a large superconducting solenoid. Beyond making such standards more economical and portable, this could allow combining a resistance standard in a single cryostat with a cryogenic current comparator and other components of the quantum metrology triangle~\cite{Keller2016,Scherer2012}. Such a combination recently achieved world-record precision for a current source~\cite{Brun-Picard2016}, but required three separate cryostats. While dilution refrigerator temperatures are currently needed to observe quantization of the Hall resistance in QAH systems, elucidation of dissipation mechanisms and materials development may point the way toward increasing practicality. Examinations of film thickness dependence~\cite{Feng2016}, alternative magnetic dopants~\cite{Chang2015,Grauer2015}, and growth techniques have begun to explore a range of possibilities, with a modulation-doped film in particular exhibiting $\rho_{yx} = 0.97$ $h/e^2$ even at 2 K~\cite{Mogi2015}. Also, theoretical proposals already exist for materials that could exhibit QAH near room temperature~\cite{Xu2013,Wu2014}. With this in mind, it is worth exploring the quantization and dissipation in existing materials to understand potential limitations.

Here, we present the most precise measurements reported to date of the Hall resistance in a QAH system, finding quantization to within a part in $10^6$ of $h/e^2$ and longitudinal resistivity under 10~m$\Omega$. Deviating from optimal conditions allows us to explore the nature of the dissipation in this system to understand ways to improve this performance further. In particular, we find that increasing the current density beyond a critical value causes a rapid rise in dissipation, echoing the breakdown phenomenon in the QH effect.

\begin{figure}
\includegraphics[width=\columnwidth]{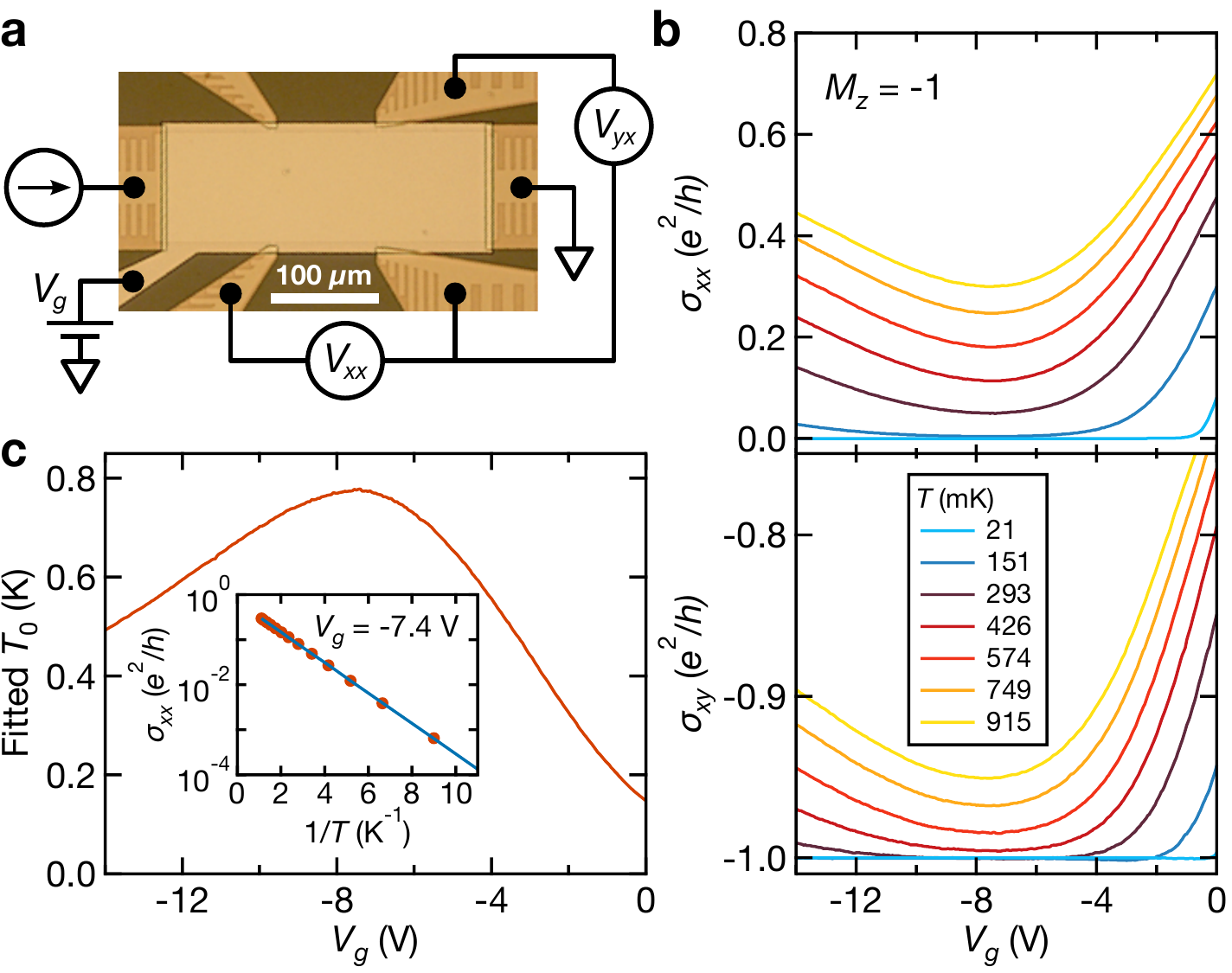}
\caption{\label{fig1}QAH device and dependence of conductivity on gate voltage and temperature. (a) Micrograph of top-gated Hall bar based on 6-nm-thick film of (Cr$_{0.12}$Bi$_{0.26}$Sb$_{0.62}$)$_2$Te$_3$. A simplified schematic of the measurement scheme is overlaid. (b) Longitudinal and transverse conductivities, $\sigma_{xx}$ and $\sigma_{xy}$, respectively, derived from lock-in amplifier measurements of the resistivity as a function of gate voltage $V_g$, for temperatures between 21 mK and 915 mK. At the lowest temperatures, the conductivities plateau over a wide range in $V_g$ at values consistent with $\sigma_{xy} = -e^2/h$ and $\sigma_{xx} \approx 0$ to within expected experimental error. (c) Fitted temperature scale $T_0$ for thermally-activated conduction, $\sigma_{xx} \propto e^{-T_0/T}$, as a function of $V_g$, peaking at 780 mK. Inset, an Arrhenius plot of $\sigma_{xx}$ as a function of $1/T$ at $V_g = -7.4$ V with a fit (blue line) to thermal activation.}
\end{figure}

With a 6-quintuple-layer sample of (Cr$_{0.12}$Bi$_{0.26}$Sb$_{0.62}$)$_2$Te$_3$ grown on a GaAs substrate by molecular beam epitaxy, we used photolithography to fabricate a Hall bar (device 1), 100 $\mu$m wide with a 100 $\mu$m center-to-center distance between 2-$\mu$m-wide voltage terminals (Fig.~\ref{fig1}(a)). The film was etched with Ar ion milling to define a mesa, and 5 nm/100 nm, Ti/Au ohmic contacts were deposited by e-beam evaporation. After growing a 40 nm Al$_2$O$_3$ dielectric over the entire sample by atomic layer deposition, we patterned and deposited a Ti/Au gate, which allows tuning of the chemical potential in the film. Finally, we removed the remaining uncovered alumina with a chemical etch before wire bonding.

\begin{figure*}
\includegraphics[width=2\columnwidth]{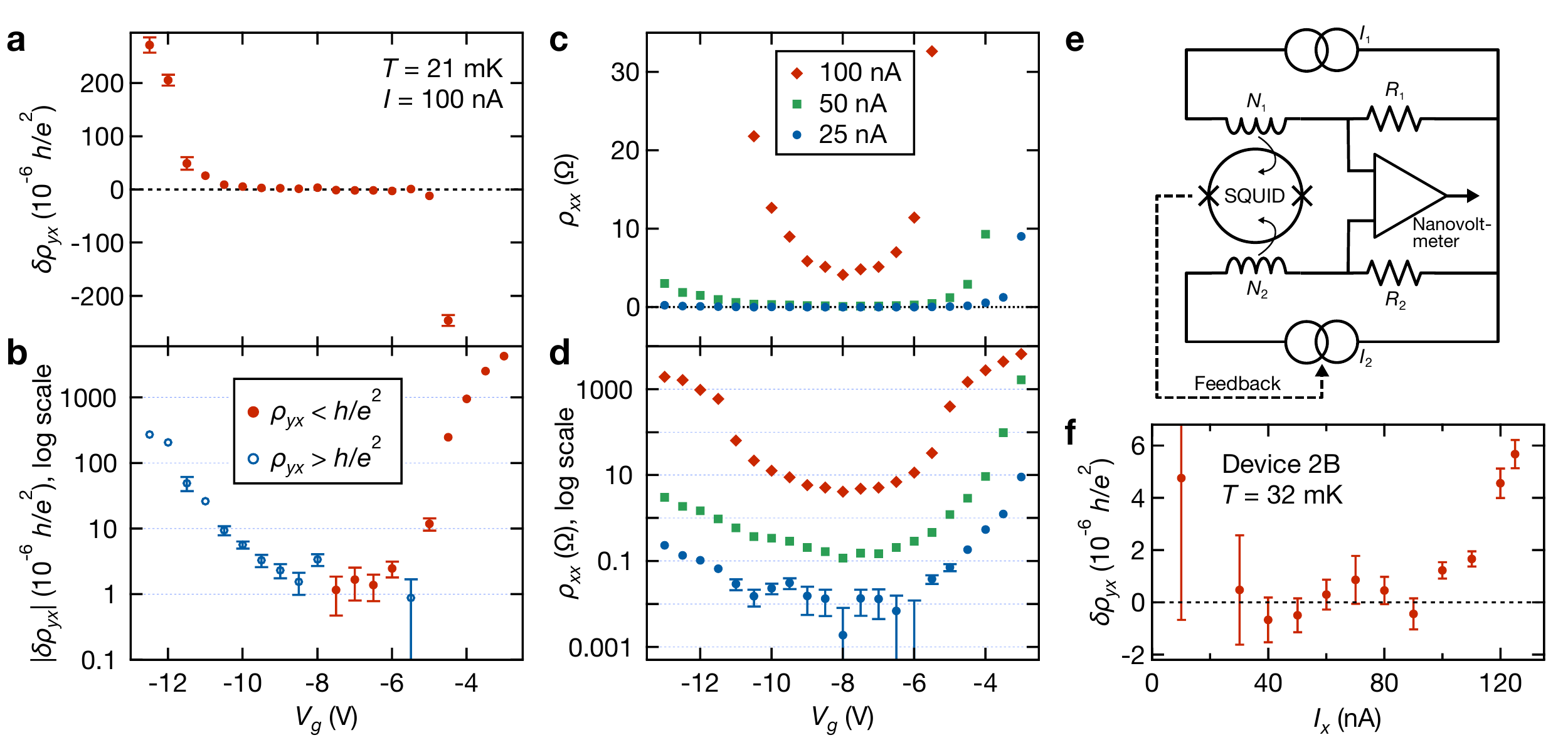}
\caption{\label{fig2}Cryogenic current comparator measurements. (a) Precise measurements of $\rho_{yx}$ using a 100 nA current at 21 mK show a plateau of $\delta \rho_{yx} = |\rho_{yx}| - h/e^2 \approx 0$ for a range of $V_g$, indicating the accuracy of quantization. In (b), the same data are plotted as $|\delta \rho_{yx}|$ on a log scale. (c,d) $\rho_{xx}$ measured as a function of $V_g$ for three different bias currents shows strong current and gate voltage dependence, displayed on linear (c) and log (d) scales. At 25 nA with $V_g$ near the center of the plateau, the resistivity is near-vanishing and measurements approach the noise floor. (e) Simplified schematic of the cryogenic current comparator. (f) Measurements of $\delta\rho_{yx}$ in a different Hall bar (device 2B) as a function of current $I_x$, showing accurate quantization for $I_x \le 90$~nA. In all plots, error bars show the standard uncertainty and are omitted when smaller than the marker.}
\end{figure*}

To observe QAH, we first measured $\rho_{xx} = V_{xx}/I$ and $\rho_{yx} = V_{yx}/I$ using standard lock-in amplifier techniques with a 5 nA bias current and the sample cooled to 21 mK in a dilution refrigerator. After magnetizing the film by applying a field $\mu_0 H = -0.5$~T and then reducing the applied field to zero, $\rho_{xx} \approx 1$~k$\Omega$ and $|\rho_{yx}| \approx 0.997$~$e^2/h$ with the gate grounded. Tuning the gate voltage $V_g$, we found a wide plateau where $\rho_{yx}$ is constant to within two parts in $10^4$ and $\rho_{xx}$ vanishes to the accuracy of our measurement, from which we infer $\rho_{yx}$ is accurately quantized at $-h/e^2$ (this is verified later). More specifically, lock-in measurements yielded $\rho_{xx} \approx 9$~$\Omega$ between the voltage terminals adjacent to the source in the clockwise direction around the edge, the direction of edge state chirality in this configuration~\cite{Bestwick2015}, and $\rho_{xx} \approx -1$~$\Omega$ on the opposite side. These nonzero values are consistent with systematic error due to leakage currents into the voltage preamplifiers~\cite{Fischer2005,Bestwick2015}, which have 100~M$\Omega$ input impedances.

By heating the sample above base temperature, we observed the onset of dissipation, shown in Fig.~\ref{fig1}(b) in terms of the longitudinal conductivity ${\sigma_{xx} = \rho_{xx}/(\rho_{xx}^2 + \rho_{yx}^2)}$ and transverse conductivity ${\sigma_{xy} = \rho_{yx}/(\rho_{xx}^2 + \rho_{yx}^2)}$. Within the plateau, the conductivity appears thermally activated, with $\sigma_{xx} \propto e^{-T_0/T}$ when larger than the systematic offsets due to leakage currents. The fitted thermal activation temperature scale $T_0$, shown in Fig.~\ref{fig1}(c), peaks at 780 mK with $V_g = -7.4$ V, indicating that we can tune the Fermi level through the apparent center of the gap. If the thermal activation fit held to our lowest temperatures, we would expect to find $\sigma_{xx} < 10^{-15}$~$e^2/h$ and correspondingly small deviations from quantization, assuming behavior similar to QH~\cite{Furlan1998}, but measuring this is far beyond the capabilities of our standard lock-in methods.

To overcome the limitations in precision of the lock-in setup, we turned to measuring with a cryogenic current comparator (CCC), a device typically used in quantum Hall metrology~\cite{Poirier2009}. A simplified schematic of a CCC is shown in Fig.~\ref{fig2}(e). With this instrument, two low-noise current sources drive currents $I_1$ and $I_2$ through two resistors $R_1$ and $R_2$ which are to be compared. The ratio of currents is precisely balanced by using windings that are inductively coupled to a superconducting quantum interference device (SQUID). The measurement of the net flux with the SQUID produces a feedback signal that keeps the net flux constant and thereby holds the ratio of the two currents fixed according to the relation $I_1 N_1 = I_2 N_2$, where $N_1$ and $N_2$ are the numbers of windings in each current loop. With the proper choice of $N_1$ and $N_2$, the voltage drops across $R_1$ and $R_2$ are approximately the same, and the difference $V = I_1 R_1 - I_2 R_2$ is measured with a nanovoltmeter. In this way, a precise measurement of the ratio of the two resistors is obtained.

Using a commercial CCC system~\cite{Drung2009,Drung2013}, we compared the Hall resistance of our magnetic TI Hall bar with a 100 $\Omega$ resistance standard calibrated at the National Institute of Standards and Technology (NIST), shipped to Stanford, and later remeasured at NIST to confirm accuracy to within one part in $10^7$. Measuring with a 100 nA current across the Hall bar, we observed a plateau in the deviation of the Hall resistance from quantization $\delta \rho_{yx} = |\rho_{yx}| - h/e^2$ as a function of $V_g$, shown on a linear scale in Fig.~\ref{fig2}(a) and log scale in Fig.~\ref{fig2}(b) (each plotted data point represents $\sim$60 individual measurements~\cite{SuppInfo}\nocite{Milliken2007,Yang2015,Efros1975,Pollak1976,Grannan1992,Liu_C2016,Ebert1983}, and error bars show the standard uncertainty). Within the plateau, for $V_g$ between -5.5 V and -9.5 V, $\rho_{yx}$ is quantized to within 3.5 ppm, whereas approaching the plateau edges $|\delta \rho_{yx}|$ grows approximately exponentially with $V_g$. Using measurements of $\rho_{xx}$ (discussed below) as a metric, we can define a ``central" region of the plateau such that $\rho_{xx}<2 \rho_{xx}^\text{min}$, which yields the interval [-9 V, -6.5 V] in $V_g$. Pooling the 332 Hall measurements made at 6 values of $V_g$ in this range, we find the average to be $\overline{\delta \rho_{yx}} = (0.67 \pm 0.28)\times 10^{-6}$~$h/e^2$, where the uncertainty reflects the standard deviation of the mean.

We would expect $\delta \rho_{yx} < 0$ in the presence of dissipation, as homogeneous bulk conduction should reduce the Hall voltage by allowing transverse current to flow across the Hall bar. Likewise, conduction via additional nonchiral edge states would reduce the measured Hall resistance~\cite{Wang2013}. Counter to this expectation, as $V_g$ was tuned to the left side of the $\rho_{yx}$ plateau we observed positive $\delta \rho_{yx}$ as the deviation from quantization increased (Fig.~\ref{fig2}(b)). Similar anomalous ``overshoot" of the quantized value for the Hall resistance can sometimes be seen in the QH effect when the current splits between multiple (evanescent) incompressible strips of different filling factors that are narrower than the Fermi wavelength~\cite{Sailer2010}, or when geometric effects lead to mixing of $\rho_{xx}$ into $\rho_{yx}$~\cite{Wel1988}. The former case has no clear analogue in QAH, where a single edge state is predicted. We speculate that this behavior results from spatially inhomogeneous bulk conduction~\cite{SuppInfo}, and would be reversed with increased dissipation (at higher temperatures, e.g., as in Fig.~\ref{fig1}(b)), though further investigation is required to clarify this point.

For an additional check of the CCC and resistance standard calibration, we measured an epitaxial graphene~\cite{Yang2017} sample designed to provide an $h/e^2$ resistance. The device consists of a pair of triple-series-connected Hall bars~\cite{Delahaye1993,Poirier2004}. When the graphene is tuned to a $\nu = 2$ quantum Hall state, a quasi-Hall voltage can be measured between voltage leads on the two Hall bars that gives a four-terminal resistance of $h/e^2$. With 1 $\mu$A across the graphene device and an applied field $\mu_0 H = -6$ T, we observed this four-terminal resistance with equivalent $\delta \rho_{yx} = (-6.5 \pm 4.6)\times 10^{-8}$~$h/e^2$, confirming the calibration of the resistance standard to well beyond the accuracy with which we measured quantization in the QAH device. To verify proper operation of the CCC at the lower currents used for QAH, we also measured the graphene device at 100 nA and found quantization with $\delta \rho_{yx} = (1.9 \pm 4.0)\times 10^{-7}$~$h/e^2$, using 120 individual measurements.

Using the CCC's nanovoltmeter, which has $\sim$100 G$\Omega$ input impedance~\cite{Drung2011}, we separately measured $\rho_{xx}$ in the magnetic TI Hall bar for the same range of $V_g$. Viewed on a linear scale in Fig.~\ref{fig2}(c), there is a clear plateau near $\rho_{xx}=0$ at gate voltages for which $\rho_{yx}$ is well-quantized, and a sharp rise as $V_g$ is tuned outside the central region. However, at 100 nA, the same current at which $\rho_{yx}$ was measured, we surprisingly found $\rho_{xx}$ still above 4 $\Omega$ at minimum. The same measurements with lower current, 50 nA and 25 nA, plotted in Fig.~\ref{fig2}(c--d), show the strong current dependence of the dissipation in this range, with $\rho_{xx}$ at 25 nA between 2 and 4 orders of magnitude lower than at 100 nA, and a minimum value of $\rho_{xx} = 1.9 \pm 6.2$ m$\Omega$ at $V_g = -8$ V.

Reducing the bias current increases the noise in Hall measurements, but can also lead to better quantization. Our best measurements of quantized Hall resistance were obtained from a separate Hall bar (device 2B, see below) at lower currents, shown in Fig.~\ref{fig2}(f). At 100~nA and above, there appear to be consistent deviations in $\rho_{yx}$ from $h/e^2$. Averaging the Hall measurements for currents ${I_x \le 90}$~nA, weighted by the inverse variance of the mean, yields $\overline{\delta \rho_{yx}} = (0.04 \pm 0.26)\times 10^{-6}$~$h/e^2$.

\begin{figure*}
\includegraphics[width=1.5\columnwidth]{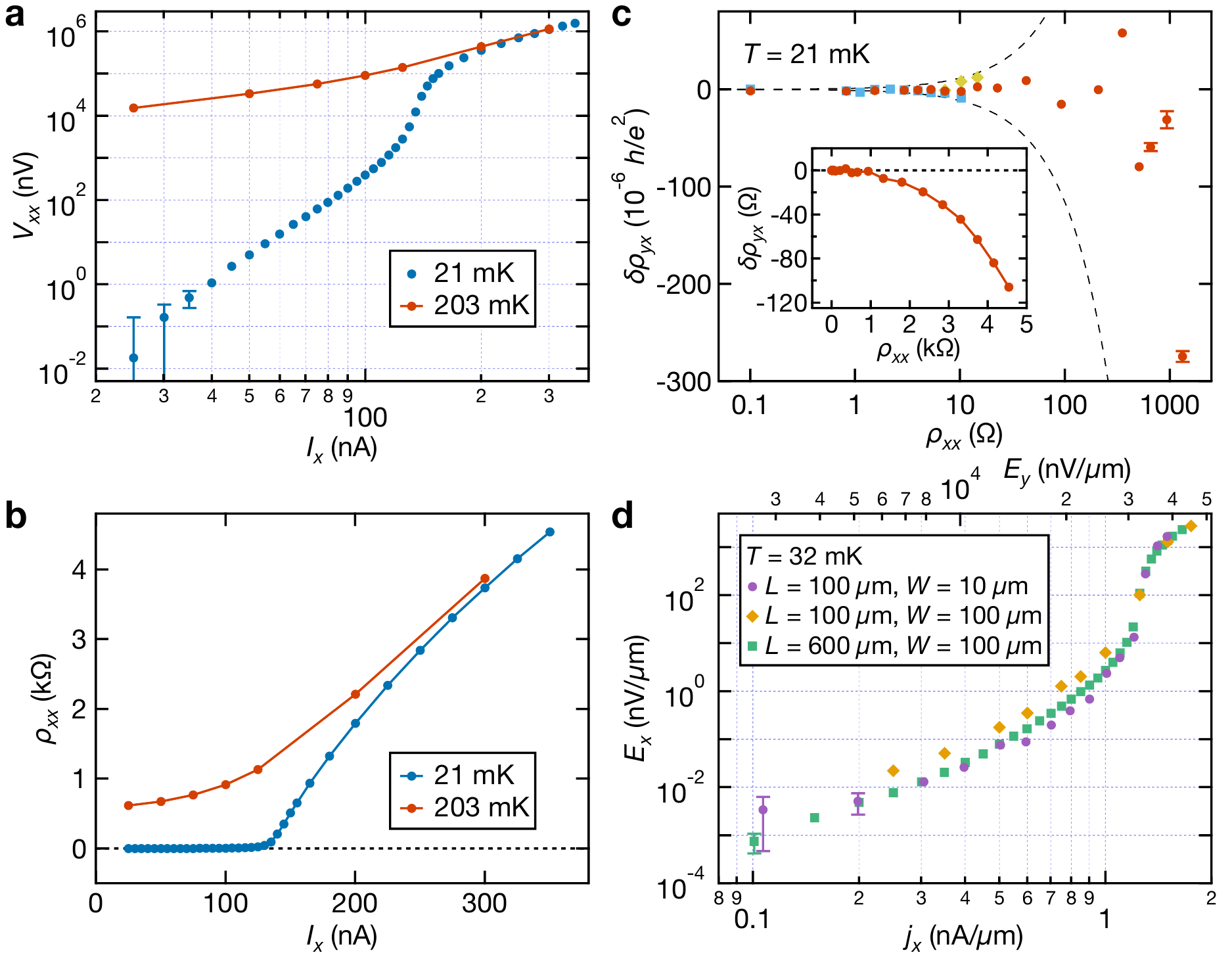}
\caption{\label{fig3}Current-induced breakdown of the QAH effect. (a) Longitudinal voltage $V_{xx}$ measured with the nanovoltmeter as a function of bias current $I_x$ with $V_g = -7.5$ V, shown for lattice temperatures of 21 mK and 203 mK as measured by the mixing chamber plate thermometer. At base temperature and $I_x \le 100$~nA, we observe an apparent power law $V_{xx}\propto I_x^{6.4}$. After a sharp rise in $V_{xx}$ for $T=21$ mK starting around $I_x = 125$ nA, the curves for the two temperatures nearly overlap, consistent with runaway electron heating. (b) $\rho_{xx}$ calculated from the data in (a), shown on a linear plot versus $I_x$. (c) Hall resistance deviations $\delta \rho_{yx}$, measured on three separate upward sweeps of current (shown with different colors and symbols---data shown in blue and yellow were taken on the opposite pair of voltage contacts from that in red), plotted against separate measurements of $\rho_{xx}$ from (b). Almost all points below $\rho_{xx}=1$ k$\Omega$, where $|\delta\rho_{yx}| < 10^{-4}e^2/h$, fall within $|\delta\rho_{yx}| < 0.03 \rho_{xx}$, shown by the dashed lines. Inset, $\delta\rho_{yx}$ exhibits quadratic dependence on $\rho_{xx}$ at higher dissipation. (d) Longitudinal electric field $E_x$ versus current density $j_x$ for three Hall bars of varying size (devices 2A-C), each at their optimum gate voltage, showing geometry-independent behavior at high current density and similar behavior even at low current density. The average transverse electric field $E_y = \rho_{yx} j_x$ shown on the top axis is calculated assuming $\rho_{yx} = h/e^2$.}
\end{figure*}

To further investigate this current dependence, we measured $V_{xx}$ for currents $I_x$ between 25 and 350 nA with the gate tuned near the center of the plateau, $V_g = -7.5$ V. The relationship between $V_{xx}$ and $I_{x}$, shown on a log-scale plot in Fig.~\ref{fig3}(a) or as $\rho_{xx} = V_{xx}/I_x$ versus $I_x$ in Fig.~\ref{fig3}(b), is markedly nonlinear. At base temperature in particular, we find an abrupt increase in longitudinal resistivity reminiscent of the current-induced breakdown of the QH effect~\cite{Nachtwei1999}. Although this rise in $\rho_{xx}$ is not as sharp as is usually observed in QH breakdown, here $\rho_{xx}$ nonetheless increases by more than a factor of 22 as the current is increased by 20\% starting at $I_x = 125$ nA.

At higher temperatures, the breakdown effect is smeared out. Data taken at 203 mK, displayed in Fig.~\ref{fig3}(a--b), indicate $\rho_{xx}$ is relatively temperature-independent at higher currents in the breakdown regime, while at lower currents the traces taken at different temperatures diverge significantly. A smoother current dependence of similar form to that seen at 203 mK was recently observed in a QAH system by Kawamura \emph{et al.}~\cite{Kawamura2017}, who attributed the behavior to electric-field-driven VRH at low temperatures. Here, the development of a sharp rise in $\rho_{xx}(I_x)$ at low temperatures suggests the bootstrap electron heating (BSEH) model of QH breakdown~\cite{Komiyama2000} may explain breakdown behavior in QAH as well.

The BSEH model ascribes the sharp increase in dissipation to runaway heating of the electron system, whose temperature $T_e$ diverges from the lattice temperature $T_L$ to settle in a steady state in which energy gain and loss rates are balanced, as given by the equation~\cite{Komiyama2000}

\begin{equation}\label{eq1}
\sigma_{xx}E_y^2 = \frac{Z(T_e) - Z(T_L)}{\tau},
\end{equation}

\noindent
where $Z(T)$ is the areal energy density of the electron system at temperature $T$ and $\tau$ is the temperature-dependent electron energy relaxation time. The transverse electric field $E_y$ is related to the current by $E_y = \rho_{yx}j_x$, and at currents relevant for breakdown, the average field $E_y \approx j_x h / e^2$. The left-hand side of Eq.~\ref{eq1} is the rate of energy gain per unit area $G = j \cdot E = \sigma_{xx} E^2 \approx \sigma_{xx} E_y^2$ since $E_y \gg E_x$, while the right-hand side gives the rate $L$ of energy loss to the lattice per area. The strong $T_e$-dependence of the conductivity enables ``bootstrap" heating of the electron system when the current is increased enough to cause this balance to become unstable, with $\partial G/\partial T_e > \partial L/\partial T_e$, until a new stable equilibrium is found at higher $T_e$. In a QH system at sufficiently low temperature, this can be observed as a discontinuous jump in $\sigma_{xx}$ with increasing $j_x$.

In our QAH measurements, we instead find a sharp but continuous rise, but this is consistent with QH experiments at higher temperatures~\cite{Komiyama1985} and the BSEH model, which predicts a continuous transition when $k_B T_L$ exceeds $\sim$6\% of the gap between Landau levels. Though the sample appears to reach $k_B T_L / \Delta \approx 0.014$ at a 21~mK base temperature, where $\Delta = 2 k_B T_0 \approx 130$~$\mu$eV is the gap extracted from thermal activation, the lack of a discontinuous jump could reflect differences in the form of $Z(T)$, with a gapped Dirac band structure in place of sharp Landau levels, and in energy relaxation processes. Additional electron heating due to electronic noise, known to generically cause $T_e$ to diverge from $T_L$ at the lowest temperatures reachable with dilution refrigerators, could also contribute to smoothing, whereas with larger gaps in QH systems it may be negligible. At higher temperatures above $\sim$10\% of the gap size, the breakdown transition is expected to be entirely smeared out according to the BSEH model, which is indeed what we observe at 203~mK where $k_B T_L/\Delta \approx 0.13$ (Fig.~\ref{fig3}(a--b)).

We additionally measured $\delta\rho_{yx}$ for varying currents $I_x$ at 21 mK. Fig.~\ref{fig3}(c) displays these data parametrically as $\delta\rho_{yx}(I_x)$ versus $\rho_{xx}(I_x)$ from Fig.~\ref{fig3}(b). In contrast to the QH effect, where $\delta\rho_{yx} \propto \rho_{xx}$ is typically seen up to $\rho_{xx} \approx 10$ $\Omega$~\cite{Jeckelmann2001,Furlan1998}, we do not observe a linear relationship, and $\delta\rho_{yx}$ measurements are not repeatable as deviations from quantization grow. At intermediate currents in particular, around the breakdown observed in $\rho_{xx}$, significant fluctuations in $\delta\rho_{yx}$ occur as the current is varied. Nonetheless, the deviations remain surprisingly small compared to $\rho_{xx}$, with $|\delta\rho_{yx}| < 8 \times 10^{-5}$~$h/e^2 \approx 2$~$\Omega$ for $\rho_{xx} < 1$ k$\Omega$. For thermally-driven deviations, on the other hand, lock-in measurements at 5 nA give a significantly larger $\delta\rho_{yx}$ of $-33$~$\Omega$ when $\rho_{xx}=704$~$\Omega$ at $T=240$~mK. A possible explanation for this behavior could be spatially inhomogeneous dissipation near breakdown, fluctuating as the current is increased, leading to enhanced $\rho_{xx}$ with lesser impact on $\rho_{yx}$~\cite{Cage1983}. As increasing current drives $\rho_{xx}$ above 1~k$\Omega$, we find a crossover to quadratic dependence $\delta\rho_{yx} \propto \rho_{xx}^2$, which has been seen previously in temperature-driven measurements in a QAH system~\cite{Bestwick2015}. We have also performed these measurements at higher temperatures, finding a faster rise in $|\delta\rho_{yx}|$ with increasing $\rho_{xx}$, and checked that longitudinal and Hall measurements do not differ substantially between different pairs of contacts~\cite{SuppInfo}.

An interesting question related to the onset of significant dissipation with breakdown is the location of current flow. Literature to date has explained QAH transport phenomena as resulting from edge channel conduction in the sense of B\"uttiker's model of the QH effect~\cite{Buttiker1988}, and one proposed source of dissipation at low temperatures is the additional presence of non-chiral edge states with the bulk still insulating~\cite{Wang2013}. Though an edge conduction model has been used successfully to describe many features of the QH effect, in certain circumstances, including at high currents relevant for metrology, experiments show that current flows predominantly in the bulk~\cite{Nachtwei1999,Jeckelmann2001} and the BSEH model discounts edge-transport effects in breakdown. Indeed, pure edge transport in the QAH effect appears inconsistent with our results. The current carried by a chiral edge channel in the B\"uttiker model is $I = (e/h)\Delta\mu$, where $\Delta\mu$ is the chemical potential difference between opposite edges. If the current flowed only at the edges, we would expect breakdown due to strong enhancement of tunneling into the bulk as $\Delta\mu$ approached the size of the gap~\cite{Nachtwei1999}. The gap extracted from the fit to thermal activation would predict an edge conduction breakdown at only $\sim$5 nA, strongly hinting that the bulk also plays a role at the higher currents measured here even in the pre-breakdown, low-dissipation regime.

For further insight on this point, we consider the geometrical scaling of this effect. For breakdown based on edge conduction alone, the breakdown current $I_{\text{cr}}$ should be relatively independent of the Hall bar width $W$ in the transverse direction. If instead bulk conduction is dominant and approximately homogeneous, we would expect $I_{\text{cr}} \propto W$, and dissipation as measured by $V_{xx}$ should scale with the distance $L$ between voltage contacts. Indeed, in the QH effect, current flow in the bulk leads to linear scaling of $I_{\text{cr}}$ with $W$ for samples that are sufficiently large compared to relevant length scales (e.g. scale of density fluctuations)~\cite{Nachtwei1999}. Thus, measuring the size dependence of breakdown behavior can provide additional clues to the nature of the current distribution.

On a separate chip from the same film growth as device 1, we fabricated three additional devices of varying size, and measured longitudinal voltage as a function of current. These Hall bars are 100 $\mu$m wide with one square between voltage terminals, 100 $\mu$m wide by 6 squares, and 10 $\mu$m wide by 10 squares (devices 2A, 2B, and 2C, respectively). Each device was measured with $V_g$ tuned to the approximate center of the resistivity plateau (respectively for devices 2A-C: -7 V, -5.8 V, -6.65 V). To account for the geometrical differences between the Hall bars, the longitudinal electric field $E_x = V_{xx}/L$ is plotted against the current density $j_x = I_x/W$ in Fig.~\ref{fig3}(d). In each case, there is a sharp increase in $E_x$ around $j_x = 1.2$ nA/$\mu$m indicating the expected linear dependence of breakdown current on width, and the behavior of $E_x (j_x)$ in the breakdown regime appears to be nearly independent of sample size. Based on this geometrical scaling, we infer that breakdown must take place through bulk conduction. We note that the strength of these conclusions is limited by the small number of samples, though Kawamura \emph{et al.}~\cite{Kawamura2017} have similarly observed that the characteristic current for crossover from low $V_{xx}$ to a higher-dissipation, linear $I$-$V$ regime is roughly proportional to sample width.

\begin{figure}
\includegraphics[width=\columnwidth]{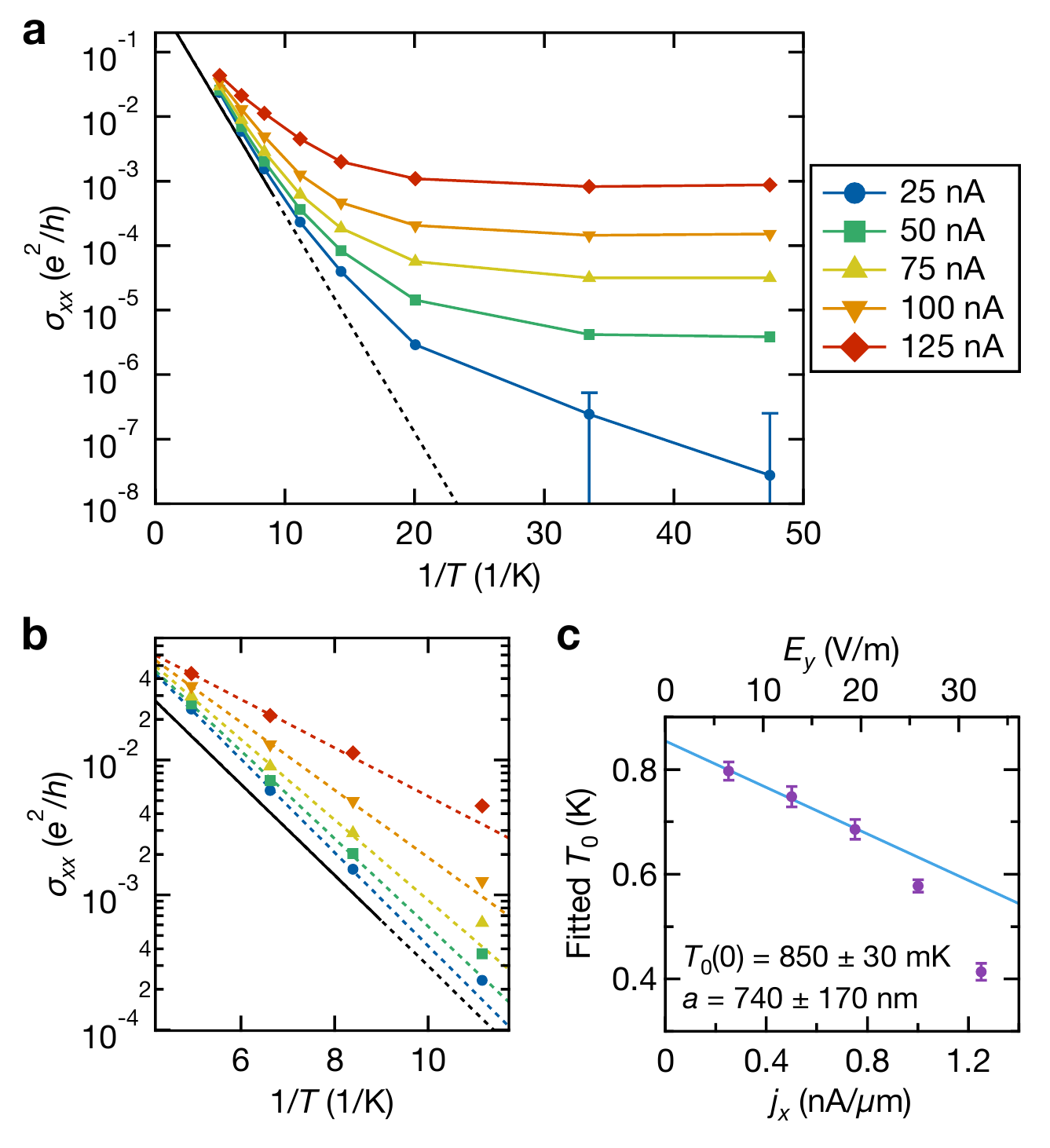}
\caption{\label{fig4}Temperature dependence of $\sigma_{xx}$ for $V_g = -7.5$ V. (a) Arrhenius plot of the temperature dependence of $\sigma_{xx}$ for five bias currents between 25 and 125 nA. The conductivity appears relatively temperature independent below 50 mK, while above 100 mK it may be thermally activated, although we cannot rule out variable-range hopping conduction based on the temperature dependence alone. The black solid line shows the fit to thermal activation for measurements with a lock-in amplifier at 5 nA AC bias current (Fig.~\ref{fig1}(c), inset), and the dashed line extrapolates the fit to lower temperatures. (b) Detail view of $\sigma_{xx}$ versus $1/T$ for higher temperatures, with fits to activated conductivity (colored dashed lines) for measurements above 100 mK. (c) Fitted activation temperature scale $T_0$ as a function of current density $j_x$ or transverse electric field $E_y \approx (h/e^2)j_x$. A fit (blue line) to $T_0 (E_y) = T_0(0)-aeE_y/k_B$ for $j_x \le 0.75$ nA/$\mu$m yields ${T_0(0) = 850 \pm 30}$~mK and ${a = 740 \pm 170}$ nm.}
\end{figure}

An explanation for behavior in the pre-breakdown regime is less clear. $E_x(j_x)$ shown in Fig.~\ref{fig3}(d) is also similar for the three Hall bars in pre-breakdown, suggesting that bulk conduction may be involved. However, it does not appear to be as uniform between the devices as in breakdown itself: $E_x$ values differ by a factor of up to 4 for similar $j_x$. Intriguingly, in the original (and lowest temperature) cooldown of device 1 we observed a power-law relationship between $V_{xx}$ and $I_x$ over nearly one order of magnitude in current and four orders of magnitude in voltage (Fig.~\ref{fig3}(a)). A fit between 25 and 100 nA yields $V_{xx} \propto I_x^{6.4}$. Power-law scaling of voltage with current is known to occur for tunneling into QH edge states~\cite{Chang1996}, but for ${\nu = 1}$, the closest analogue to the QAH state, both theory and experiment find $V \propto I$. In the absence of bulk conduction, Luttinger liquid behavior due to the possible presence of additional, quasi-helical edge states theoretically predicted in some QAH systems could perhaps lead to a different power-law scaling. Nonetheless, our evidence for current flow in the bulk suggests this explanation is unlikely. Moreover, in contrast to the power law in device 1, measurements of devices 2A-C at a slightly higher temperature show clear curvature in the log-scale plot of $E_{x}$ versus $j_x$ (Fig.~\ref{fig3}(d)), and the computed $\sigma_{xx}$ appears exponential in $j_x$ below breakdown~\cite{SuppInfo}.

Measurements of $V_{xx}$ as a function of current and temperature up to the onset of breakdown in device 1 provide additional hints about the pre-breakdown behavior. In Fig.~\ref{fig4}(a), we display these data as $\sigma_{xx}$ in an Arrhenius plot for $I_x$ between 25 and 125 nA. For $T \le 50$~mK, $\sigma_{xx}$ is nearly temperature independent, particularly for higher currents. At least two explanations are plausible. First, the electron temperature may be sufficiently elevated that varying the lattice temperature has little effect in this range. In the BSEH model, this would correspond to $Z(T_L) \ll Z(T_e)$ in Eq.~\ref{eq1} for $T_L$ under $\sim$50~mK and sufficiently large current. Second, the behavior of $\sigma_{xx}$ at low temperature may be dominated by hopping transport driven by the Hall electric field (recalling that $E_y = \rho_{yx} j_x \approx j_x h / e^2$). In QH systems, non-ohmic conduction at the lowest temperatures and high but subcritical currents has been ascribed to field-driven VRH of the form $\sigma_{xx}(I_x) = \sigma_{xx}^I \exp\left(-\sqrt{E_1/E_y}\right)$, where $E_1$ is a characteristic field and the prefactor $\sigma_{xx}^I$ is only weakly dependent on temperature and field~\cite{Furlan1998,Jeckelmann2001}. Field-driven VRH has also been proposed~\cite{Kawamura2017} as an explanation of the current dependence of $\sigma_{xx}$ in the QAH effect, and a fit of our low-temperature data to such a model is plausible~\cite{SuppInfo}. Whether or not temperature-independent hopping is responsible for part of the pre-breakdown behavior, however, the presence of the breakdown effect indicates that a crossover must take place to a regime in which electron heating is the dominant effect.

At elevated temperatures, $\sigma_{xx}$ clearly has a strong temperature dependence. Above 100~mK, higher current DC measurements can be plausibly fit to thermal activation, shown in a detail view in Fig.~\ref{fig4}(b), but transport is still clearly non-ohmic. Although for each current alone, the temperature dependence of the conductivity in this range can be reasonably fit to thermally driven VRH with ${\sigma_{xx} \propto \exp\left(-(T_1/T)^{1/(d+1)}\right)}$ for $d=1$, 2, or 3, the typical model for non-ohmic VRH does not clearly agree with the data~\cite{SuppInfo}. Instead, the similar form ${\sigma_{xx} \propto \exp(-T_0/T)}$ that matches both lock-in and DC measurements suggests activated conduction is more likely. In Fig.~\ref{fig4}(c), we plot the fitted thermal activation temperature scales from these data against $j_x$ and $E_y$, showing the reduction in $T_0$ with increasing $E_y$.

In pre-breakdown QH, a linear reduction in the activation energy with Hall electric field has frequently been observed and suggested to be caused by either a field-dependent broadening of the Landau level extended state bands~\cite{Shimada1998} or the tilted potential over the length scale of localized state wavefunctions reducing the energy required for excitation to extended states~\cite{Komiyama1985}. Supposing this latter explanation may apply in a QAH system for excitations from midgap localized states at the Fermi level to the surface state bands, we fit the data for lower currents, $j_x \le 0.75$~nA/$\mu$m, to ${T_0 (E_y) = T_0(0) - aeE_y/k_B}$. The fit yields ${T_0(0) = 850 \pm 30}$~mK, somewhat larger than the 780~mK scale found in lock-in measurements, but in reasonable agreement, and ${a = 740 \pm 170}$~nm. This value of $a$ is fairly consistent with that extracted from a fit of data at fixed temperature: in devices 2A-C, ${\sigma_{xx} \propto \exp(a e E_y / k_B T)}$, with $a \approx 600$~nm, for $j_x \le 1$~nA/$\mu$m~\cite{SuppInfo}. Although $T_0(E_y)$ departs from this dependence at higher fields, the nonlinearity could be due to electron heating near the onset of breakdown causing the thermal activation fit to be inaccurate.

Based on our interpretation, the temperature and current dependence of the conductivity we have observed in the pre-breakdown regime appears to result from an interplay of electron heating and dissipation in the bulk. This dissipation seems to occur through electric-field-assisted thermal activation to the surface state bands and possibly also field-dependent VRH, both of which relate to localization. As other authors~\cite{Li2016,Kawamura2017} have suggested, a better understanding of localization and the origin of midgap states in QAH systems will be important in the pursuit of more robust quantization.

In summary, we have demonstrated that the QAH effect presents a promising platform for resistance metrology with accurate quantization of the Hall resistance to within one part per million at zero magnetic field. The primary limitation to the precise measurement of quantization at low temperatures is a current-induced breakdown of the low-dissipation state, which we attribute to runaway electron heating in bulk current flow. Though currently available materials show strong temperature dependence, and the small gap leads to breakdown at significantly smaller currents than in QH samples, new material systems~\cite{Xu2013,Wu2014} and improved growth techniques~\cite{Mogi2015} to increase the exchange-induced gap, reduce disorder in the gap size, or lower the density of midgap states may make QAH metrology more practical. With existing materials, our results suggest Hall bar arrays or simply wider Hall bars will allow for higher current and therefore lower noise measurements of quantization. Further study may also elucidate the complicated pre-breakdown behavior of the conductivity we have observed here.


\begin{acknowledgments}
We thank Marlin Kraft for performing resistance standard calibrations at NIST, and Marc Kastner, Biao Lian, and Shoucheng Zhang for helpful discussions. Device fabrication, measurements, and analysis were supported by the U.S.\ Department of Energy, Office of Science, Basic Energy Sciences, Materials Sciences and Engineering Division, under Contract DE-AC02-76SF00515. Infrastructure and cryostat support were funded in part by the Gordon and Betty Moore Foundation through Grant GBMF3429. X.~K., L.~P.\ and K.~L.~W.\ acknowledge support from FAME, one of six centers of STARnet, a Semiconductor Research Corporation program sponsored by MARCO and DARPA, and from the Army Research Office under Grants W911NF-16-1-0472 and W911NF-15-1-0561:P00001. X.~K.\ acknowledges support from the Chinese National Thousand Young Talents Program and Shanghai Sailing Program under Grant No.\ 17YF1429200. Part of this work was performed at the Stanford Nano Shared Facilities (SNSF), supported by the National Science Foundation under Award ECCS-1542152.

Identification of commercial products or services used in this work does not imply endorsement by the U.S.\ government, nor does it imply that these products are the best available for the applications described.
\end{acknowledgments}


\bibliography{QAHE-metrology}


\balancecolsandclearpage


\section{Supplemental Material}

\stepcounter{mysection}
\setcounter{section}{1}
\setcounter{equation}{0}
\setcounter{figure}{0}
\renewcommand{\theequation}{S\arabic{equation}}
\renewcommand{\thefigure}{S\arabic{figure}}
\renewcommand{\citenumfont}[1]{S#1}

\subsection{Methods}

\subsubsection{Film growth}
All of the magnetic topological insulator Hall bars used in this work were fabricated from the same film, grown by the same molecular beam epitaxy (MBE) method as those in Refs.~\citeS{S_Kou2014} and~\citeS{S_Bestwick2015}. A 6-quintuple-layer, high-quality single-crystalline Cr-doped (Bi$_\text{x}$Sb$_\text{1-x}$)$_\text{2}$Te$_\text{3}$ film was grown in an ultra-high-vacuum Perkin-Elmer MBE system. A semi-insulating GaAs (111)B substrate was loaded into the growth chamber and annealed to 580 $^\circ$C in a Se-rich environment to remove the native oxide. During the growth, the GaAs substrate temperature was maintained at 200 $^\circ$C, with the Bi, Sb, Te, and Cr shutters opened at the same time. Epitaxial growth was monitored by \emph{in situ} reflection high-energy electron diffraction (RHEED). After the film growth, 2 nm Al was evaporated \emph{in situ} to passivate the surface at room temperature and then exposed to air to form an oxide layer, protecting the film from unwanted environmental doping or other possible aging effects.

\subsubsection{Fabrication}
The Hall bars fabricated from the magnetic TI film were patterned using contact photolithography. For each patterning step, the sample was spin coated with a hexamethyldisilazane adhesion layer followed by Megaposit SPR 3612 photoresist. The pre-exposure bake at 80 $^\circ$C for 120 s was chosen to minimize possible damage to the film from exposure to high temperatures. The photoresist was exposed under an ultraviolet mercury vapor lamp with a dose of approximately $70$ mJ/cm$^2$ and then developed in Microposit CD-30 for 35 s. Stripping the resist after etching or lift-off after metallization was performed with a standard acetone/isopropanol rinse.

The device mesas were defined by etching the surrounding film with Ar ion milling. The sections of the mesa underneath the ohmic contacts were patterned into comb-like shapes, which can be seen in Fig.~\ref{fig1}(a) of the main text. We have empirically found such patterns tend to reduce the contact resistance compared to shapes of similar area but smaller perimeter. The ohmic contacts, consisting of a 5 nm Ti sticking layer and 100 nm Au, were deposited by e-beam evaporation following a brief \emph{in situ} Ar ion etch to further improve the contact resistance. For electrostatic gating, a dielectric layer was first grown uniformly across the film by evaporating a 1 nm Al seed layer, allowing it to oxidize, and then depositing approximately 40 nm of alumina by atomic layer deposition using trimethylaluminum and water precursors. The top gate was then patterned and deposited by evaporating 5 nm Ti and 85 nm Au. Excess alumina dielectric on the surrounding area was etched using the tetramethylammonium-hydroxide-based developer Microposit MF CD-26.

\subsubsection{Lock-in measurements}
Initial current-biased measurements were performed using standard lock-in techniques at frequencies between 1 and 10 Hz. Higher frequencies result in significant phase differences between the measured signals and the excitation. Using a 1 G$\Omega$ bias resistor and 5 V RMS excitation, a 5 nA AC bias was applied to the Hall bar and measured after flowing across the device with an Ithaco 1211 current preamplifier, set to have a gain of $10^7$ V/A and input impedance of 200 $\Omega$. The differential voltages $V_{xx}$ and $V_{yx}$ were simultaneously measured with NF Corporation LI-75A voltage preamplifiers, which have a gain of 100, and 100 M$\Omega$ input impedances. The outputs of the preamplifiers were measured with Stanford Research Systems SR830 lock-in amplifiers. A voltage was applied to the gate electrode using a Yokogawa 7651 DC source.

To obtain low electron temperatures in the measured devices, measurement lines in the dilution refrigerator used for this experiment have electronic filtering at the mixing chamber stage. GHz frequencies are filtered by passing the lines through a cured mixture of bronze powder and epoxy~\citeS{S_Milliken2007}, while the MHz range is filtered with five-pole low-pass RC filters, which are mounted on sapphire plates for thermal anchoring.

\subsubsection{CCC measurements}
Precise Hall measurements were conducted by comparing to a 100 $\Omega$ resistance standard with a commercial CCC system~\citeS{S_Drung2009,S_Drung2013} using 2065-turn and 8-turn windings for the standard and the QAH sample, respectively. The actual resistance of the standard, calibrated at NIST, is 100.00018 $\Omega$ with a relative uncertainty of $\sim$$10^{-7}$, and after the experiment was conducted at Stanford, it was remeasured at NIST to confirm its value. Measurements of the resistance ratio were made by sampling the nanovoltmeter reading over 10-second cycles, during which the current direction was reversed half way through to remove DC offsets. Data points plotted in the figures generally consist of measurements over 60 cycles, although spurious outlying samples (typically 0-2 per set of 60) were removed in some cases.

Longitudinal resistivity was measured with the nanovoltmeter~\citeS{S_Drung2011} of the CCC system, which has an input impedance of $\sim$100 G$\Omega$, by turning off and shorting across the current source in the current loop with the resistance standard. With the connections to the QAH sample properly configured for a longitudinal measurement, the nanovoltmeter then directly measures $V_{xx}$, and $\rho_{xx} = V_{xx}/I$ was obtained by dividing by the applied current. As with Hall measurements, data points in the figures typically consist of measurements over 60 cycles.

\subsubsection{Graphene standard}
Graphene was grown on SiC with the Si-terminated side facing a graphite disk in a graphite-lined furnace filled with Ar. The substrate was first processed in forming gas at 1050 $^\circ$C for one hour, and then was annealed at 1900 $^\circ$C for 120 s. Hall bars 100 $\mu$m wide and 600 $\mu$m long were fabricated on a single chip using a clean photolithography process~\citeS{S_Yang2015}, which starts with deposition of a thin Pd/Au layer on the as-grown graphene to prevent contamination from the further fabrication steps. The metal protecting the graphene in the Hall bar area was etched by diluted aqua regia at the last step. The aqua regia used for metal etching introduces $p$-doping in the graphene, which is intrinsically $n$-type due to transfer of charge from the substrate, so that the carrier density of the samples is usually very close to the Dirac point. The carrier density was then tuned to the optimal level by a low temperature annealing process~\citeS{S_Yang2017}. Finally, the graphene devices were encapsulated with 720 nm Parylene C to help maintain the carrier density level during the transfer from NIST to Stanford in the ambient air.

Two Hall bars with similar performance in preliminary measurements were selected for the composite array and connected with gold wire bonds. Characterization at 1.6~K and between -6~T and -9~T indicated the devices were well-quantized at tens of $\mu$A. The array of two Hall bars with a triple-series connection produces a quasi-Hall voltage $V_H = 2 I \rho_{yx}$ between contacts on the two devices, though only for one direction of the magnetic field. When the graphene is in a $\nu=2$ quantum Hall state, the array resistance standard produces a four-terminal resistance of $h/e^2$.

\subsubsection{Error and statistical methods}
Error bars in plots of data obtained with CCC system reflect the standard uncertainty (standard deviation of the mean), and are omitted when the uncertainty is smaller than the marker size. Curve fitting was performed using nonlinear least squares, weighted using the standard uncertainty of each data point when applicable, and reported errors are the estimated standard deviations of the fit coefficients.

For most fits in both the main text and the supplement, assuming that the standard uncertainty accurately estimates the true measurement error results in large values of $\chi^2$ compared to the number of degrees of freedom, suggesting that either there is an unrecognized source of systematic error or the model used for fitting is incomplete. The latter explanation is likely most important: the measurements we fit to thermally activated and VRH conductivity have been made in a regime where there may be competing effects, such as electron heating, and the possibly complicated and spatially varying density of states could cause deviations from these simple models. We contend that these models may nonetheless be useful approximations, as explained elsewhere in the main text and supplement. Systematic errors may also explain some of the discrepancy. However, a divergence between the lattice temperature measured by the thermometer and the electron temperature of the sample (when electron heating due to current flow is negligible), which is common to measurements in dilution refrigerators and only significant at the lowest temperatures, would not account for the deviations from the fits that we observe. In particular, this phenomenon cannot explain discrepancies between measurements at fixed temperature and the models we apply, and the fits of temperature dependence in Fig.~\ref{fig4} of the main text and Fig.~\ref{figS_intermedVRH} of the supplement are performed over temperature ranges likely too high for the effect to be significant (measurements reported in Ref.~\citeS{S_Bestwick2015} using the same cryostat and electronic filtering setup suggest electron temperatures lower than 30~mK can be attained in this system).

Because of these issues, standard calculations for least squares fitting may underestimate the uncertainty of fit coefficients for our approximate models when the measurement error is assumed to be purely statistical. To avoid this, we instead compute the uncertainties of the parameters by assuming the model is a good fit to the data (that is, the true measurement error is taken to be proportional to the standard uncertainty of each data point but large enough that the fit is reasonable).

\subsection{A model for Hall resistance larger than the quantized value}

\begin{figure}
\includegraphics[width=\columnwidth]{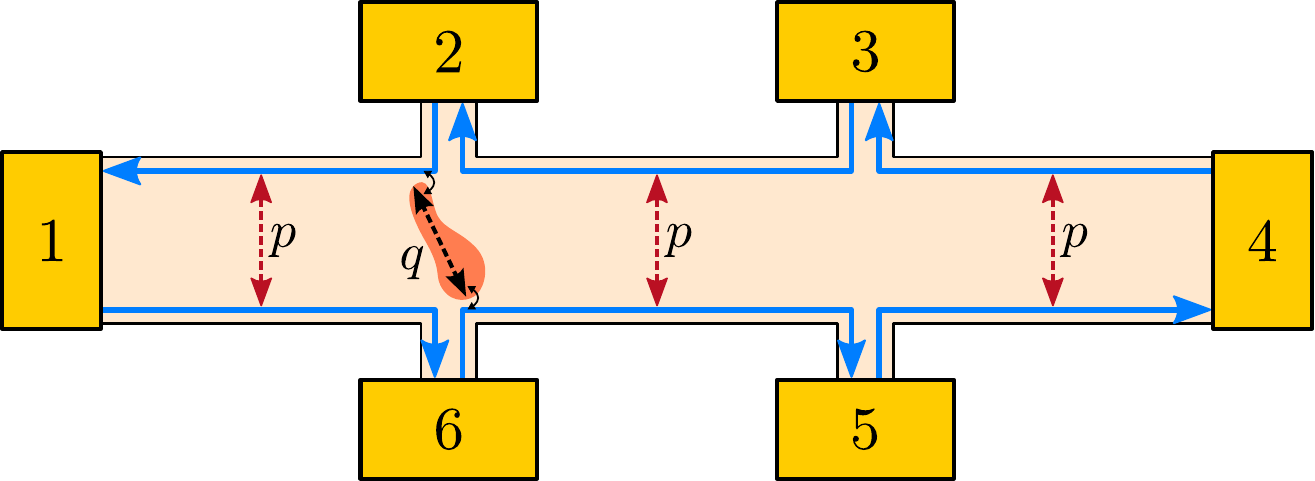}
\caption{\label{figS_hallbar}Simplistic transport model resulting in measuring $\delta\rho_{yx} > 0$. QAH edge states propagate with counter-clockwise chirality. Electrons may scatter across the Hall bar from one edge to another. Inhomogeneous bulk conductivity results in a conductive puddle coupling the edge channel between terminals 1 and 2 to that between 5 and 6.}
\end{figure}

In Fig.~\ref{fig2}(b--c) of the main text, $\delta\rho_{yx} = \left|\rho_{yx}\right|-h/e^2$ is surprisingly positive for $V_g$ on the left side of the plateau. We would usually expect to observe $\delta\rho_{yx} \le 0$ since nonvanishing, homogeneous sheet conductivity should decrease the measured Hall resistance due to shorting through the bulk, and the presence of additional, non-chiral edge states would also reduce the Hall voltage. Below, we propose a simplistic model for how the measured Hall resistance could exceed $h/e^2$ due to spatially inhomogeneous bulk conduction in combination with edge conduction. We do not claim that this model is realistic, but provide it to illustrate one possible way this situation could occur.

Suppose a six-terminal Hall bar exhibits the QAH effect with edge channel conduction having counter-clockwise chirality. We consider two sources of dissipation other than the inevitable equilibration between the 1D mode and the contacts. First, we assume a finite probability of scattering across the Hall bar to the opposite edge, denoted by $p$. For simplicity, we take this probability to be the same between each pair of opposite edges. Second, suppose there is a conductive puddle in the bulk that couples to two of the edges near opposite voltage terminals as shown in Fig.~\ref{figS_hallbar}, such that an electron in an edge channel incident on the coupling region has a probability $q$ of being diverted across the Hall bar (and crossing, longitudinally, between the two voltage terminals). Again, for simplicity, we consider coupling at isolated points directly adjacent to the voltage terminals, and assume that the presence of this conductive puddle does not affect the inter-edge scattering $p$. We next calculate the transmission coefficients $T_{ij}$ (for transmission from terminal $j$ to terminal $i$) in order to find the measured four-terminal resistances through the Landauer-B\"uttiker formalism~\citeS{S_Buttiker1988}.

Consider, for example, an electron originating from terminal 6 in the edge channel heading towards terminal 5. With probability $q$, the electron passes through the bulk puddle to the edge between terminals 1 and 2; with probability $1-p$, it then continues on to terminal 1, rather than scattering across the Hall bar again and returning to terminal 6. Thus, we see $T_{16} = q(1-p)$. Alternatively, with probability $1-q$, the electron continues along the edge towards terminal 5, and thereafter scatters across to the edge between terminals 2 and 3 with probability $p$. Hence, $T_{26} = (1-q)p$ and $T_{56} = (1-q)(1-p)$ (and $T_{36} = T_{46} = 0$). Continuing this reasoning, we find the nonzero off-diagonal transmission coefficients to be:
\begin{subequations}
	\begin{align}
		T_{53} = T_{35} &= p, \label{eq:right_T}\\
		T_{26} = T_{62} &= (1-q)p, \label{eq:left_T}\\
		T_{16} = T_{52} &= q(1-p), \\
		T_{56} = T_{12} &= (1-q)(1-p) \\
		T_{61} = T_{23} &= T_{34} = T_{45} = 1-p.
	\end{align}
\end{subequations}
Note that the diagonal (reflection) coefficients $T_{ii}$ will not enter into subsequent calculations.

Given the transmission coefficients, currents and voltages are related by the equation
\begin{equation}\label{eq:landauer}
	I_i = \frac{e^2}{h} \sum_{j} \left(T_{ji} V_i - T_{ij} V_j \right),
\end{equation}
where $I_i$ is the current from terminal $i$ into the Hall bar and $V_i$ is the voltage at terminal $i$. We can convert Eq.~\ref{eq:landauer} to the form $I_i = \sum_j G_{ij} V_j$, defining the conductance matrix
\begin{equation}
G_{ij} = \frac{e^2}{h} \left( \delta_{ij}\sum_k T_{ki} - T_{ij}\right),
\end{equation}
where $\delta_{ij}$ is the Kronecker delta. Shifting all voltages by a constant does not change the current, so $G$ is singular. However, choosing a terminal $k$ and setting $V_k = 0$, the submatrix $\widetilde{G}$ of $G$ with the $k$-th row and column removed is invertible. Knowing the currents into the device, we can then solve for the voltages using $V_i = \sum_{j \neq k} \widetilde{G}^{-1}_{ij} I_j$ for $i \neq k$.

For current-biased Hall measurements, we take $I_1 = -I_4 = I_\text{sd}$ at the current contacts, and $I_2 = I_3 = I_5 = I_6 = 0$ for voltage terminals. We define a set of four-terminal resistances as follows: Hall resistances $R_{yx}^\text{L} = V_{62}/I_\text{sd}$ and $R_{yx}^\text{R} = V_{53}/I_\text{sd}$ for the left and right pairs of voltage contacts, respectively, and longitudinal resistances $R_{xx}^\text{T} = V_{23}/I_\text{sd}$ and $R_{xx}^\text{B} = V_{65}/I_\text{sd}$ for the top and bottom, respectively. Solving as described above, we find
\begin{subequations}
	\begin{align}
		R_{yx}^\text{L} &= \frac{h}{e^2} \left( \frac{1}{1-q} \right) \approx \frac{h}{e^2} \left( 1+ q \right), \\
		R_{yx}^\text{R} &= \frac{h}{e^2}, \\
		R_{xx}^\text{T} &= \frac{h}{e^2} \left( \frac{p}{1-p} \right) \approx \frac{h}{e^2} p, \\
		R_{xx}^\text{B} &= \frac{h}{e^2} \left( \frac{p}{1-p} + \frac{q}{1-q} \right) \approx \frac{h}{e^2} \left( p+q \right),
	\end{align}
\end{subequations}
expanding to first order in $p$ and $q$ (assuming $p,q \ll 1$) on the right-hand side. Evidently, if the measured value of $\rho_{yx}$ is taken to be $R_{yx}^\text{L}$, then $\delta\rho_{yx} / (h/e^2) = q > 0$.

In this model, the Hall resistance is always $h/e^2$ or larger, but clearly we do not expect such behavior in a real device. Indeed, lock-in measurements at elevated temperatures (Fig.~\ref{fig1} of the main text) show $|\rho_{yx}| < h/e^2$ and decreasing as dissipation increases. We can make a simple, if naive, modification to the model to show that we can reproduce this behavior. Suppose there is also a homogeneous component to the bulk conductivity, which allows transport through the bulk between the voltage contacts in the transverse direction (we treat this bulk conduction in parallel and neglect any effect on longitudinal transport, since accounting for it will not qualitatively change our conclusions). We take the new transmission coefficients, $T'_{ij}$, to be related to the old $T_{ij}$ in Eqs.~\ref{eq:right_T} and \ref{eq:left_T} by ${T'_{ij} = T_{ij} + s}$ for ${(i,j) \in \{(2,6),(6,2),(3,5),(5,3)\}}$. In other words, we add an additional component $s$ to the transmission between opposite pairs of voltage terminals. To first order in $p$, $q$, and $s$, we find
\begin{subequations}
	\begin{align}
		R_{yx}^\text{L} &\approx \frac{h}{e^2} \left( 1 + q - s\right), \\
		R_{yx}^\text{R} &\approx \frac{h}{e^2} \left( 1 - s \right), \\
		R_{xx}^\text{T} &\approx \frac{h}{e^2} \left( p+s\right), \\
		R_{xx}^\text{B} &\approx \frac{h}{e^2} \left( p+q+s \right).
	\end{align}
\end{subequations}
As before, $R_{yx}^\text{L}$ increases with $q$, but now both Hall resistances decrease as $s$ grows. We could imagine a scenario where $q>s$ and $q$ grows faster than $s$ as $V_g$ is tuned away from the center of the $\rho_{yx}$ plateau, but increasing the temperature eventually leads to $s > q$ and $R_{yx}^\text{L,R} < h/e^2$. In the case of homogeneous conductivity $(q=0)$, we recover a result we would expect: ${R_{yx}^\text{L} = R_{yx}^\text{R} < h/e^2}$ and $R_{xx}^\text{T} = R_{xx}^\text{B} > 0$.

Though this model may not apply in the high-current regime explored in this work, where we suggest there may be significant current in the bulk, inhomogeneous conductivity could also lead to mixing of $\rho_{xx}$ into $\rho_{yx}$ through oblique current flow in such a scenario~\citeS{S_Wel1988}, with similar results. Hence, for either edge-dominated or bulk-dominated conduction, spatially inhomogeneous bulk conductivity could lead to measuring $\delta\rho_{yx} > 0$.

\subsection{Additional longitudinal and Hall resistivity measurements vs.\ current}

\begin{figure*}
\includegraphics[width=0.9\textwidth]{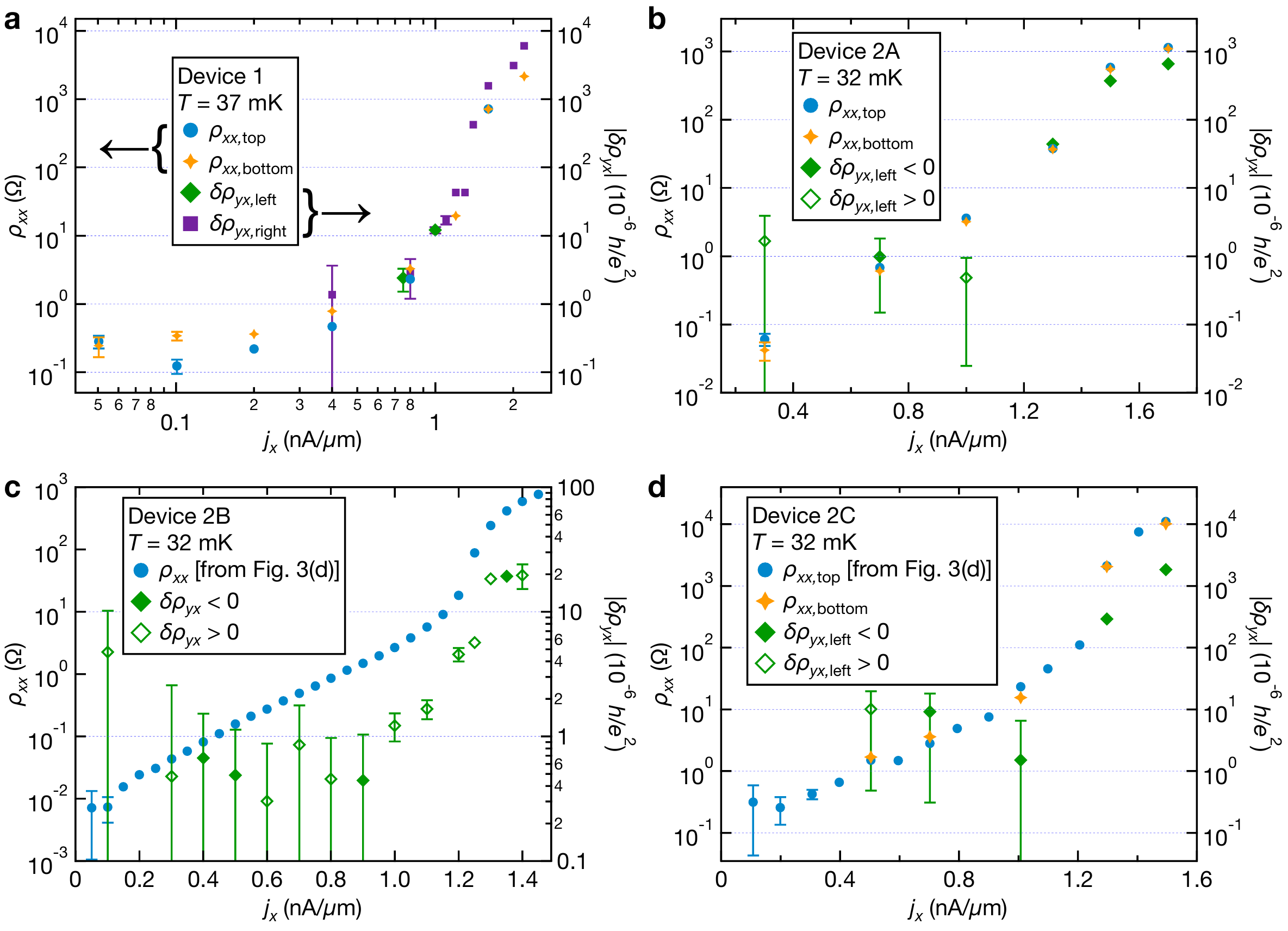}
\caption{\label{figS_addl}Additional measurements of Hall and longitudinal resistivities. (a) $\rho_{xx}$ and $\delta\rho_{yx}$ measured between all pairs of contacts on a second cooldown of device 1 at 37 mK. Data were taken for each pair on a separate upward current sweep. For all of these Hall measurements, $\delta\rho_{yx}<0$. (b) Additional measurements of resitivities of device 2A as a function of current density, taken on a single upward current sweep rotating between the two longitudinal contact pairs and one Hall pair. (c) $\delta\rho_{yx}$ vs. $j_x$ in device 2B, measured on a separate upward current sweep. These measurements are also partially shown in Fig.~\ref{fig2}(f) of the main text. For reference, we also display $\rho_{xx}$ calculated from the data in Fig.~\ref{fig3}(d) of the main text. (d) $\rho_{xx}$ measured on the opposite pair of contacts from those used for measurements shown in Fig.~\ref{fig3}(d) (also plotted here) along with $\delta\rho_{yx}$ measurements from device 2C.}
\end{figure*}

In addition to the data shown in the main text, we made further measurements of $\rho_{xx}$ and $\rho_{yx}$ as a function of current, displayed in Fig.~\ref{figS_addl}. All of these measurements were taken at the respective optimal gate voltages of the devices, determined by finding the value of $V_g$ that minimizes $\rho_{xx}$ in lock-in measurements taken at elevated temperatures (where a minimum can be distinguished), as in the main text. In QH metrology, measurements of longitudinal and Hall voltages can vary between different sides of a Hall bar or different pairs of contacts in some devices~\citeS{S_Jeckelmann2001}. To check whether this could be occurring in our devices, we measured across different contact pairs in three of the devices. Some differences can be seen between $\rho_{xx}$ measurements across the two pairs of contacts at lower currents, which could be due to spatial inhomogeneities in the resistivity, but they remain well within an order of magnitude of each other and agree quite well at higher currents.

\subsection{Fits to variable-range hopping conduction}

\begin{figure*}
\includegraphics[width=0.92\textwidth]{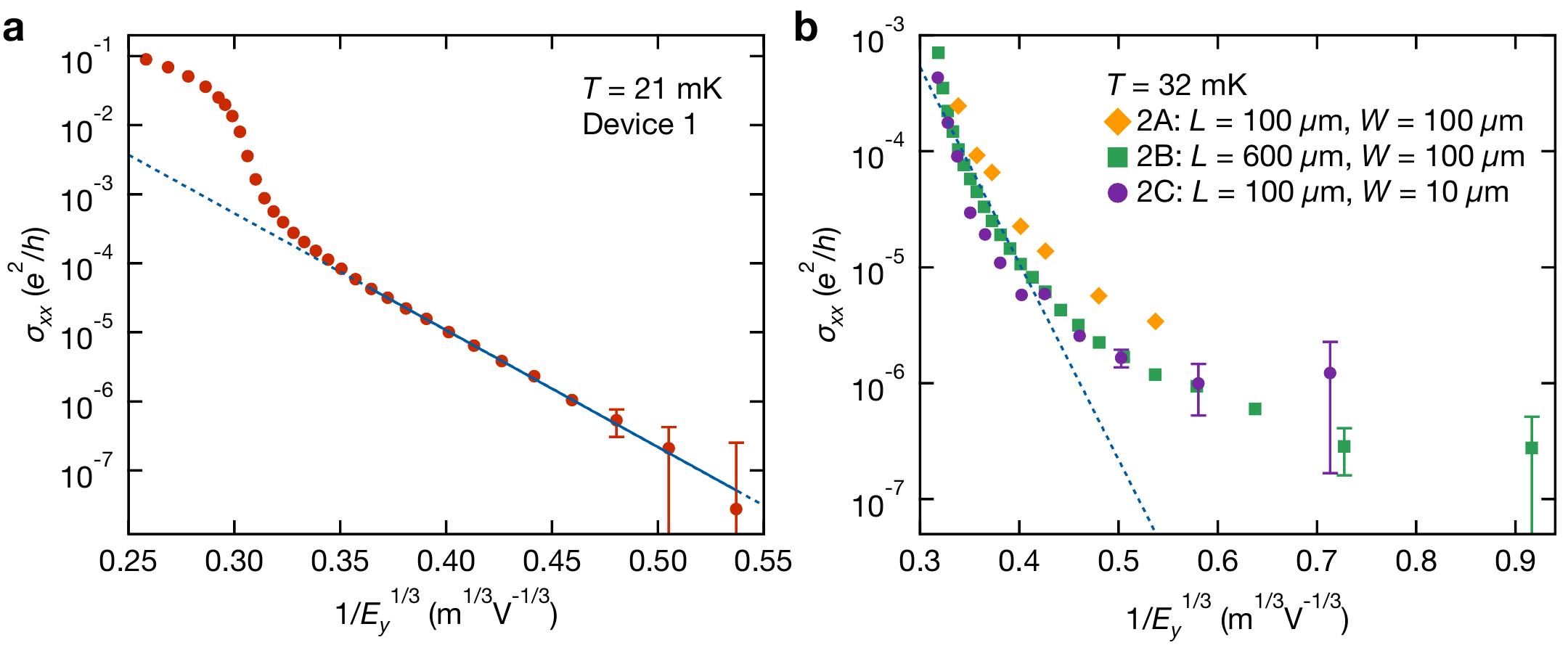}
\caption{\label{figS_fieldVRH}Possible field-driven hopping conductivity. (a) Longitudinal conductivity $\sigma_{xx}$ of device 1 from the measurements shown in Fig.~\ref{fig3}(a--b) of the main text is plotted against $1/E_y^{1/3},$ where ${E_y = \rho_{yx} j_x}$ is the average transverse field and we assume ${\rho_{yx}=h/e^2}$. A fit to the temperature-independent hopping model of Eq.~\ref{eq:fieldVRH} with $\alpha = 1/3$ for $E_y \le 21$~V/m yields yields ${\sigma_{xx}^I = 63.3 \pm 0.2}$~$e^2/h$ and ${E_1 = 59 \pm 1}$~kV/m and is shown by the blue solid line. At larger fields (where the extrapolated fit is shown with a dashed line), $\sigma_{xx}$ begins to deviate from this relation, perhaps because of electron heating in combination with an additional, temperature-dependent conduction mechanism. (b) $\sigma_{xx}$ versus $1/E_y^{1/3}$ for devices 2A-C at ${T = 32}$~mK. Field-driven hopping is not a clear fit to these data over any range of $E_y$. The dashed line shows the fit for device 1 from (a) for comparison.}
\end{figure*}

As discussed in the main text, the origin of the non-ohmic behavior in the pre-breakdown QAH regime is somewhat unclear. Some form of variable-range hopping (VRH) due to localized states within the exchange-induced gap may play a role in the nearly temperature-independent conductivity at the lowest temperatures, and could provide an alternative explanation of the higher temperature behavior we suggested could be due to thermal activation.

Typically, ohmic VRH conductivity follows the relation
\begin{equation}\label{eq:tempVRH}
\sigma_{xx}(T) = \sigma_{xx}^T e^{-(T_1/T)^\alpha},
\end{equation}
where $\sigma_{xx}^T$ is a prefactor that is only weakly dependent on temperature compared to the exponential factor, $T_1$ is a characteristic temperature scale, and $\alpha=1/(d+1)$ with $d$ the number of spatial dimensions. In cases where Coulomb interactions between localized electrons suppress the density of states around the Fermi level, frequently the situation in QH systems~\citeS{S_Jeckelmann2001}, the exponent is instead predicted~\citeS{S_Efros1975} to be $\alpha=1/2$, which has been reported in some experiments~\citeS{S_Furlan1998}.

In sufficiently strong electric fields, the hopping transport becomes independent of temperature and depends only on the field. At the lowest temperatures in QH measurements, and with currents that are large but below breakdown, finite longitudinal conductivity has sometimes been attributed~\citeS{S_Furlan1998} to electric-field-driven VRH according to
\begin{equation}\label{eq:fieldVRH}
\sigma_{xx}(I_x) = \sigma_{xx}^I e^{-(E_1/E_y)^\alpha},
\end{equation}
where $E_1$ is a characteristic field scale and the prefactor $\sigma_{xx}^I$ is only weakly dependent on temperature and field. Here, $E_y = \rho_{yx} j_x = \rho_{yx} I_x / W$ is the average transverse electric field.

At intermediate fields where $\sigma_{xx}$ depends on both temperature and current, neither Eq.~\ref{eq:tempVRH} nor Eq.~\ref{eq:fieldVRH} applies. Instead, hopping conductivity in this intermediate regime is predicted~\citeS{S_Pollak1976} to follow
\begin{equation}\label{eq:PR-VRH}
\sigma_{xx}(I_x,T) = \sigma_{xx}^T \exp \left[ \frac{e E_y a}{k_B T} - \left(\frac{T_1}{T}\right)^\alpha\right],
\end{equation}
where $a$ is a length scale related to the characteristic hopping distance. This model has been successfully applied in other systems outside of the QH regime, such as doped germanium~\citeS{S_Grannan1992} and epitaxial graphene~\citeS{S_Liu_C2016}, but to our knowledge it has not been extensively explored in QH experiments. Despite the exponential current dependence of $\sigma_{xx}$ sometimes seen in pre-breakdown QH~\citeS{S_Ebert1983,S_Komiyama1985,S_Shimada1998}, the temperature dependence observed in these studies does not fit Eq.~\ref{eq:PR-VRH}. Nonetheless, below we evaluate the possibility of hopping in each of these three regimes as an explanation for the behavior we observed in pre-breakdown QAH.

At the lowest temperatures ($T\le50$~mK), we found $\sigma_{xx}$ in device 1 to be temperature independent (Fig.~\ref{fig4} of the main text). To check if the field-driven hopping model can explain the current dependence, we plot $\sigma_{xx}$ extracted from data in Fig.~\ref{fig3}(a--b) of the main text in log-scale against $1/E_y^{1/3}$ in Fig.~\ref{figS_fieldVRH}(a). A relation of the form in Eq.~\ref{eq:fieldVRH} should appear as a straight line in such a plot (up to small deviations due to the prefactor). We have chosen to show a fit for the exponent $\alpha = 1/3$ from the 2D model here, but we note that we cannot distinguish a clear best fit between $\alpha = 1/2, 1/3$ and $1/4$. In each case, the current dependence of $\sigma_{xx}$ can be reasonably fit to a field-driven VRH model for $E_y \le 21$~V/m, corresponding to a current of 80~nA (a somewhat smaller range than for the power-law relation between voltage and current that holds up to $I_x = 100$~nA, as described in the main text). Above this current, $\sigma_{xx}$ clearly deviates from the low-field fitting, with significant curvature in the log-scale plot around breakdown. Indeed, the presence of the breakdown effect, which we ascribe to electron heating, indicates that the behavior of $\sigma_{xx}$ cannot be described purely by temperature-independent hopping. Deviations from the fit to Eq.~\ref{eq:fieldVRH} could be due to a crossover to activated conductivity at higher electron temperatures. However, it is also possible that the goodness of the VRH fit at lower currents could be coincidental, an artifact of activated conduction combined with the nonlinear current dependence of the electron temperature. Measurements of devices 2A-C at somewhat higher temperature (${T=32}$~mK) do not appear to be consistent with this field-driven hopping model over any range of $E_y$, as seen in Fig.~\ref{figS_fieldVRH}(b), further suggesting either that the agreement with the model in device 1 is coincidental, or that this form of hopping can only be observed at the very lowest temperatures in these samples.

\begin{figure}
\includegraphics[width=0.95\columnwidth]{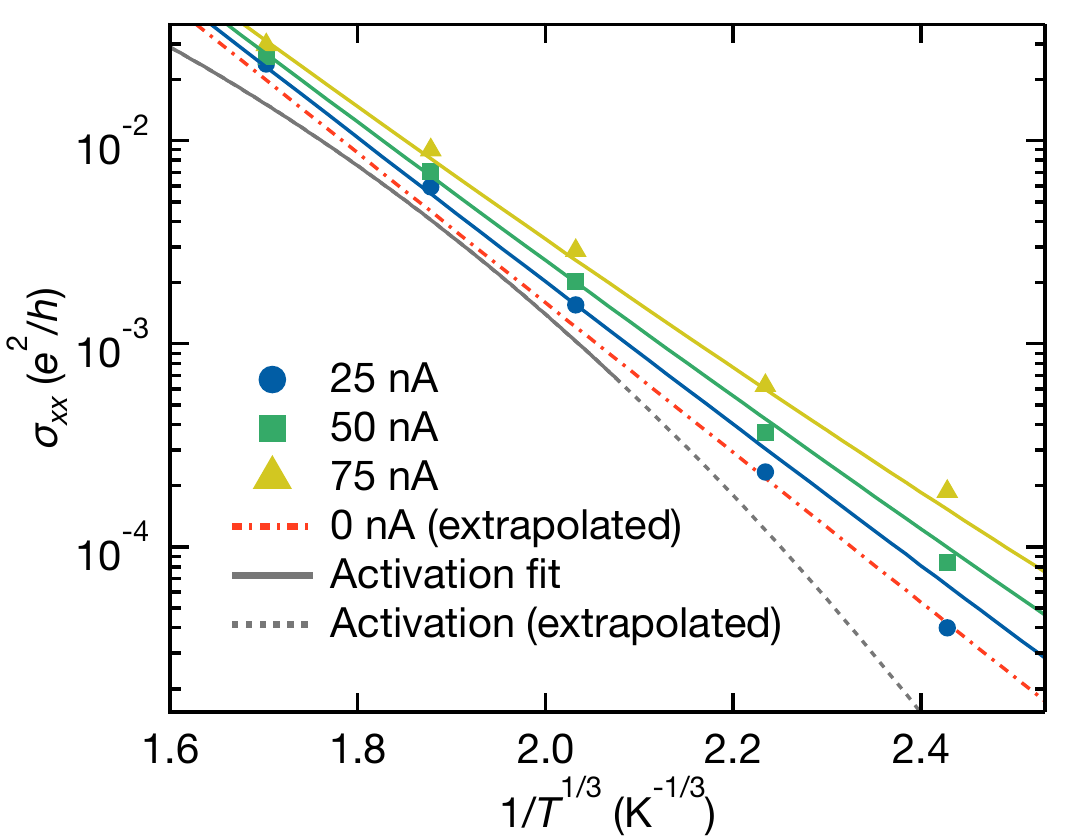}
\caption{\label{figS_intermedVRH}Fitting to a field- and temperature-dependent hopping model. Longitudinal conductivity $\sigma_{xx}$ is plotted against $1/T^{1/3}$ for ${I_x = 25}$~nA, 50 nA, and 75 nA. A least-squares fit of all of these measurements to Eq.~\ref{eq:PR-VRH} yields the solid curves with colors corresponding to the currents, with fit parameters ${\sigma_{xx}^T = (3.8 \pm 0.9) \times 10^4}$~$e^2/h$, ${a = 400 \pm 50}$~nm, and ${T_1 = 610 \pm 30}$~K. The red dash-dotted line shows an extrapolation of this fit to $I_x = 0$, the expected behavior in the ohmic regime of the VRH model. However, the ohmic value of $\sigma_{xx}$ predicted from the fit is larger (for all $T$) than lock-in measurements of $\sigma_{xx}$ at 5~nA AC bias, shown as the solid gray curve and extrapolated to lower temperatures with a dashed line.}
\end{figure}

In the higher temperature regime, VRH may also play a role in the temperature dependence of the conductivity. With $\sigma_{xx}$ data from Fig.~\ref{fig4} of the main text on a log-scale plot against $1/T^{1/3}$ in Fig.~\ref{figS_intermedVRH}, we see that the temperature dependence is reasonably close to that in Eq.~\ref{eq:tempVRH} in a range between 70~mK and 203~mK at each fixed current. However, Eq.~\ref{eq:tempVRH} cannot explain the current dependence, which is still clearly non-ohmic. A model for the behavior of $\sigma_{xx}$ in this regime must account for both current and temperature. In the main text, we suggested it could be due to electric-field-assisted thermal activation, but the field- and phonon-assisted hopping model of Eq.~\ref{eq:PR-VRH} provides another possibility. We fit $\sigma_{xx}(I_x,T)$ within this temperature range and for ${I_x \le 75}$~nA to Eq.~\ref{eq:PR-VRH}, shown in Fig.~\ref{figS_intermedVRH}, finding a length scale ${a = 400 \pm 50}$~nm. This model appears plausible, but its agreement with the data is far from perfect, particularly at lower temperatures. Moreover, we find the extrapolated ohmic VRH conductivity $\sigma_{xx}^\text{fit} (I_x=0,T)$ to be larger than the lock-in measurements of $\sigma_{xx}$ at all temperatures, contrary to our expectations (though we have not accounted for temperature dependence of the prefactor, which tends to decrease as the temperature rises~\citeS{S_Jeckelmann2001}). For these reasons, we suspect field-assisted thermal activation is more likely than VRH, but a full explanation of the behavior in the pre-breakdown regime may involve a complex interplay of all of these effects, including hopping and activated conduction along with electron heating. In any case, clarifying these origins will require further investigation.

\subsection{Exponential current dependence of conductivity in devices 2A-C}

\begin{figure}
\includegraphics[width=0.95\columnwidth]{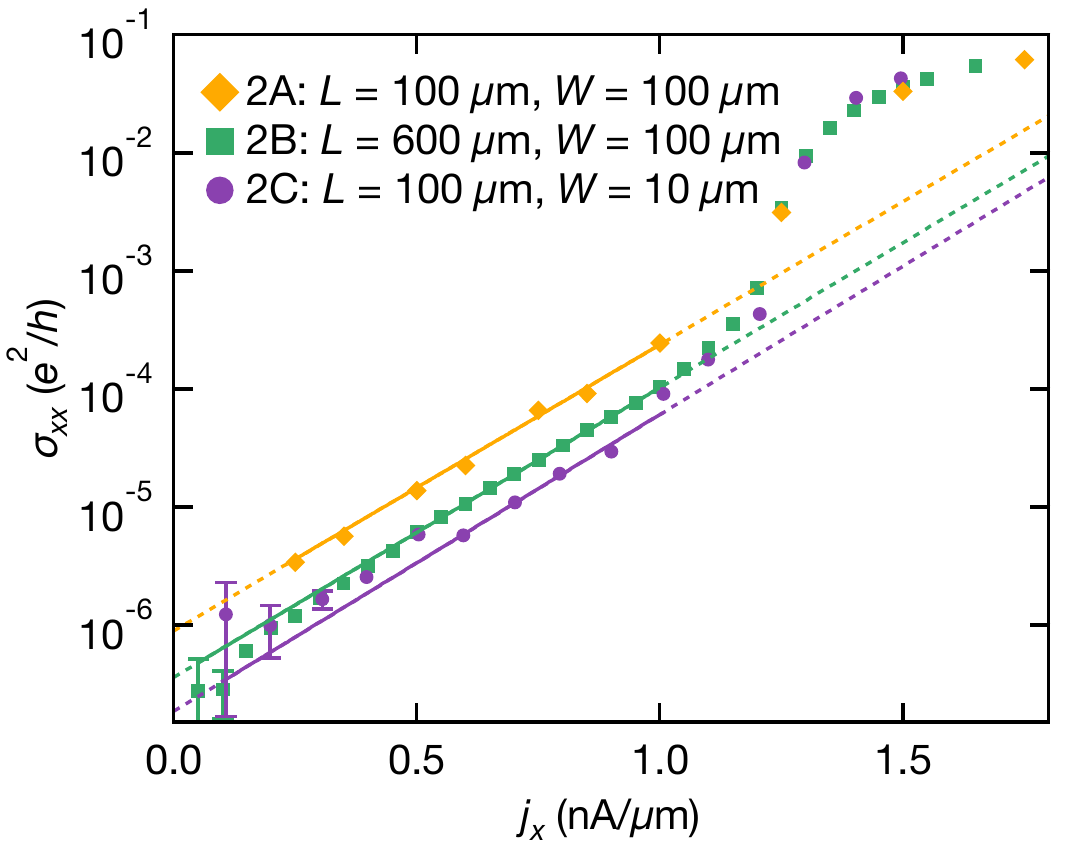}
\caption{\label{figS_dev2exp}Exponential current dependence of longitudinal conductivity. Measurements of $\sigma_{xx}$ of devices 2A-C at $T=32$~mK are plotted against $j_x$. Fits to $\sigma_{xx}=\sigma_0 \exp(aeE_y/k_B T)$ for $j_x \le 1.01$~nA/$\mu$m (pre-breakdown) are shown by the solid lines, and extrapolated with the dashed lines. Fitting yields the parameters ${a = 600 \pm 40}$~nm and ${\sigma_0 = (9 \pm 2)\times 10^{-7}}$~$e^2/h$ for device 2A, ${a = 600 \pm 30}$~nm and ${\sigma_0 = (3.6 \pm 0.1) \times 10^{-7}}$~$e^2/h$ for device 2B, and ${a = 620 \pm 60}$~nm and ${\sigma_0 = (1.9 \pm 0.7) \times 10^{-7}}$~$e^2/h$ for device 2C.}
\end{figure}

In the main text, we noted that the longitudinal conductivity of devices 2A-C in the pre-breakdown regime appears exponential in $E_y$ (or $I_x$). This behavior could be consistent with either the field-assisted thermal activation model or the hopping model of Eq.~\ref{eq:PR-VRH}. In Fig.~\ref{figS_dev2exp}, $\sigma_{xx}$, extracted from the data shown in Fig.~\ref{fig3}(d) of the main text, is plotted on a log-scale against $E_y$. Fitting to ${\sigma_{xx} = \sigma_0 \exp(aeE_y/k_B T)}$, we find similar values of the parameter $a$ for each Hall bar: ${600 \pm 40}$~nm, ${600 \pm 30}$~nm, and ${620 \pm 60}$~nm for devices 2A, 2B, and 2C, respectively. The fitted prefactor $\sigma_0$, on the other hand, differs by roughly a factor of 5 between 2A and 2C. The reason for this discrepancy is unclear, but we speculate that it might be due to inhomogeneities in the gap size and electrostatic effects effectively constricting the conduction channel differently in each device, so that the true current density differs from our estimate. These effects could have a lesser impact on the length scale $a$, which we believe is related to localization length or hopping distance.

\bibliographystyleS{apsrev4-1}
\bibliographyS{QAHE-metrology}

\end{document}